\documentclass[preprint]{aastex}
\usepackage{epsf}
\slugcomment{To appear in ApJ, accepted Jun 8, 2000}
\newcommand{\MOLH} {\hbox{${\rm H}_2$}}         
\newcommand{\HVIB} {\hbox{${\rm H}_2^*$}}         
\newcommand{\HCOP} {\hbox{{\rm HCO}$^{+}$}}     
\newcommand{\CNOA} {\hbox{C91$\alpha$}}         
\newcommand{\CSFA} {\hbox{C65$\alpha$}}         
\newcommand{\kms}{km s$^{-1}$}
\newcommand{\mic} {$\mu\hbox{m}$}
\newcommand{\percc} {$\hbox{{\rm cm}}^{-3}$}    
\newcommand{\cmsq}  {$\hbox{{\rm cm}}^{-2}$}    
\newcommand{\TKIN} {$T_{\rm kin}$}      
\newcommand{\HII}{H{\sc ii}}
\newcommand{\CII}{C{\sc ii}}
\newcommand{\solmass}{$\hbox{M}_\odot$}
\begin{document}
%
%
   \title{The location of the dense and ionized gas in the NGC~2023 PDR}

   \author{F. Wyrowski   \altaffilmark{1}, 
           C.M. Walmsley \altaffilmark{2}, 
           W.M. Goss        \altaffilmark{3}
           and A.G.G.M.~Tielens \altaffilmark{4} }

\altaffiltext{1}{Department of Astronomy, University of Maryland,
                 College Park, MD 20742-2421}
\altaffiltext{2}{Osservatorio Astrofisico di Arcetri, 
                 Largo E. Fermi 5, I-50125 Firenze, Italy}
\altaffiltext{3}{NRAO, Very Large Array, Very Long Baseline Array,
                 P.O. Box 0, Socorro, NM 87801}
\altaffiltext{4}{Kapteyn Astronomical Institute, Groningen}
%
%
   \begin{abstract} 
     The VLA and the BIMA array were used to obtain high resolution
     (10--20\arcsec) observations of C$^+$, traced by the \CNOA\ 
     recombination line at 8.6~GHz, and the dense molecular gas,
     traced by HCN and \HCOP(1--0), of the photon dominated region
     (PDR) associated with the reflection nebula NGC~2023. Using the
     VLA, continuum emission is detected at 8.6~GHz from a faint \HII\ 
     region associated with HD~37903. The \CNOA\ emission originates
     from a 0.4~pc long filament, extending from the east to the south
     of the exciting star HD~37903.  Within the filament three \CNOA\ 
     clumps can be distinguished, each associated with filamentary
     vibrationally excited \MOLH\ emission in the direction toward
     HD~37903. The \HCOP\ emission has a clumpy appearance
     superimposed on a more extended component.  \CNOA\ is, in
     general, closer to the exciting star than \HCOP\ emission as
     expected from PDR models.  The morphologies of \HCOP \ and HCN
     are quite similar. Based on the \CNOA \ linewidth towards one of
     the clumps a limit of 170~K on the kinetic temperature in the
     ionized carbon layer can be derived. This value is consistent
     with PDR models with \MOLH\ densities of about $10^5$ \percc.
     However, this result suggests surprisingly low limits on the
     turbulence in the PDR.  We detected a compact 3~mm continuum
     source in the PDR, which appears to be a cold ``core'' of density
     $10^7$ \percc , 0.03 parsec diameter, and 6 \solmass .  We
     conclude that it may have formed within the PDR.
     
     In an appendix, observations of the \CNOA\ 
     recombination line toward five additional PDRs using the Effelsberg
     100m telescope are described.
   \end{abstract}
%
 
 \section{Introduction}
 Reflection nebulae are gas and dust clouds in the vicinity of hot
 young stars. On their surfaces, photon dominated regions (PDRs) exist
 due to the interaction of the newborn stars with their parental
 molecular environment leading to intense emission of fine structure
 lines of carbon and oxygen and of H$_2$ ro-vibrational transitions
 (\HVIB). A considerable fraction of the mass of the ISM is included
 in PDRs.  They are also responsible for most of the FIR radiation in
 the Galaxy. Thus, our understanding of PDRs is a key astrophysical
 problem. Within a PDR, ionized carbon plays an important role in the
 molecular chemistry in the hot HI zone and H/H$_2$ transition layer
 \citep{ste95}. Ionized carbon has mainly been traced to date using
 the carbon $^2$P$_{3/2}$-$^2$P$_{1/2}$ fine structure line at
 158~\mic .  However, observations of this transition are limited
 presently to resolutions of about 1 arc minute, corresponding to 0.15
 parsec at the distance of the Orion nebula.  There is, however,
 another way to approach the problem of determining the spatial
 distribution of ionized carbon.  Radio observations of the carbon
 recombination lines can be used to analyze ionized carbon in PDRs
 \citep{nat94}. In this case the higher density regions of the PDR are
 selectively observed since the radio line intensity is proportional
 to the carbon emission measure.

 This procedure has already been successfully carried out towards the
 PDR of the Orion Bar \citep{wyr97a}, leading to a view of the \CII\ 
 region with the highest spatial resolution to date (0.04~pc).  A
 layer of carbon line emission bracketed between ionized and molecular
 media was detected. In contrast to predictions of current PDR models,
 the layer of ionized carbon is essentially coincident with H$_2^*$
 emission. Simple theory predicts that the radio recombination line
 should peak farther from the ionization front than molecular
 hydrogen. Thus, a possible discrepancy between observation and theory
 may exist. Hence, in order to obtain a more general knowledge of the
 properties of ionized carbon (\CII) regions, investigations of
 further PDRs with different physical conditions are needed.

 Here observations of the PDR in the reflection nebula NGC~2023 are
 presented, located at a distance of $\approx 475$~pc in the Orion
 L~1630 molecular cloud and illuminated by the B1.5 V star HD~37903.
 This star is much less luminous than the exciting stars of the Orion
 nebula and the physical conditions are expected to be different.
 NGC~2023 has been the target of many observational studies including
 [\CII] 158~$\mu$m \citep{how91,jaf94}, H$_2^*$
 \citep{gat87,fie94,fie98,mcc99,mar99}, and CN and HCN \citep{fue95}.
 In a recent study, the C91$\alpha$ radio recombination line using the
 Effelsberg 100-m telescope towards this source was observed
 \citep{wyr97b} suggesting densities of $\sim 10^5$~\percc.  Wyrowski
 et al.\ discuss several geometrical dependent models which help to
 constrain the physical parameters of the \CII\ region by comparison
 with the other PDR tracers. Face- and edge-on models of the PDR were
 analyzed as well as clumpy models.  A clear conclusion from this
 study was that improved angular resolution in the C91$\alpha$ line
 was needed; thus the VLA was used to obtain improved resolution and
 further test the models.

 In order to interpret the carbon line data, it is also useful to
 obtain molecular line observations with similarly high angular
 resolution. We therefore decided to study the stratification between
 different PDR layers (H$_2^*$, \CII, molecules) with the aid of high
 angular resolution data for the molecular phase of this edge-on PDR.
 For this purpose, the high density tracers \HCOP\ and HCN were
 observed, expected to be abundant in layers where the influence of
 the ultraviolet radiation field is still appreciable. Here we report
 BIMA observations of these molecules with an angular resolution of 10
 arc seconds, matching the VLA observations of the carbon line.

 \section{Observations}
  \subsection{Carbon recombination line measurements with the VLA}
  \label{sec-vla}

  NGC~2023 was observed with the VLA in the D array on Dec 8, 11 and
  12, 1997. Table 1 summarizes the basic observing parameters.  The
  phase center of the image was chosen to lie 1 arc minute south of
  HD~37903 ($\alpha(2000)=05^h 41^m 38\fs39$, $\delta(2000)=-02\degr
  15\arcmin 32\farcs$5) 3C84 was observed for 13 minutes each day
  and was used for bandpass calibration. Phase calibrations were made
  on QSO~0607--085 every 25 minutes. The flux density scale was
  established by observations of 3C147, assuming a flux density of
  4.74~Jy for this source.
  
  At the \CNOA\ observing frequency of 8.589104~GHz \citep{roh96} the
  correlator was used with a bandwidth of 1.5625 MHz.  Hanning
  smoothing was applied during the data acquisition and the final
  spectral resolution is 6.1~kHz (0.2 \kms ) with a total of 255
  channels.

  The data were processed using the AIPS package.  The resultant
  synthesized beam is 11.3\arcsec$\times$9.3\arcsec\ (PA=--3\degr)
  with natural weighting and the final rms noise was 1~mJy/beam
  (0.16~K) in a channel image, consistent with the theoretical noise
  level. To increase the temperature sensitivity, also lower
  resolution images were obtained from tapered UV data which have
  synthesized beams of 19.6\arcsec$\times$18.1\arcsec\ (PA=1\degr)
  resulting in an rms noise of 1.2~mJy/beam (0.06~K).  A 3.5~cm
  continuum image was produced by summing over the line free channels
  and the resulting rms is 0.1~mJy/beam.

  Due to missing short spacings the largest angular scale imaged with
  the VLA is three arc minutes. From comparison of the integrated
  observed flux density with the single dish 100m results of
  \cite{wyr97b} we estimate that the VLA observations detect 55\% of
  the peak \CNOA \ flux density and 35\% of the velocity integrated
  flux density (Fig.~\ref{missing}).  To recover the missing flux
  density of the VLA observations, an additional image of the
  integrated \CNOA\ intensity was constructed using the AIPS maximum
  entropy deconvolution task VTESS and the total flux density was
  forced to be consistent with the 100m telescope data (an average
  over all positions observed by \cite{wyr97b}).

  \subsection{\CSFA \ observation with the 100-m telescope} 

  Additionally, an observation of the \CSFA\ carbon recombination line
  was obtained with the Effelsberg 100m telescope toward one position
  ((0'',--80'') relative to HD~37903) of NGC~2023 on Dec 2, 1997.
  Calibration was made relative to NGC~7027, assuming a flux density
  of 5.5~Jy for this source at the observing frequency of
  23.4159609~GHz.  The angular resolution of the telescope at this
  frequency is 40\arcsec.

  \subsection{HCO$^+$ and HCN observations with BIMA}

  The BIMA array \citep{wel96} was used in its C and D configuration
  for four tracks on NGC~2023 between May and August 1999 obtaining
  baselines from 6 to 85~m. A 3 pointing mosaic was observed, separated
  by 67\arcsec, covering the range of the \CNOA\ emission.  Every 20
  min the phase calibrator QSO~0609--157 was observed.  Jupiter was used
  to establish the flux density calibration leading to flux densities
  between 5 and 7.5~Jy for the potentially variable quasar QSO~0609--157 on
  different days.  \HCOP\ and HCN were observed in the lower sideband
  with a spectral resolution of 0.1 MHz (0.3 \kms) and 256 spectral
  channels each.  All calibration and image deconvolution was carried
  out using the MIRIAD package: averaged line emission images were
  constructed using a robustness parameter of 0 leading to a beam size
  of 11.2\arcsec$\times$8.1\arcsec\ (PA=--7\degr). To increase the
  temperature sensitivity, line cubes were also made with natural
  weighting leading to a beam size of 15.8\arcsec$\times$10.7\arcsec\ 
  (PA=--6\degr).  All the images were deconvolved using a maximum
  entropy algorithm (MIRIAD task mosmem).  The rms noise in the
  channel images is $\sim 0.2$~Jy/beam, close to the theoretical noise
  level. Since the single dish HCN observations of \cite{fue95} cover
  parts of our image, it was possible to estimate for a few positions the
  missing flux density of the interferometer data and found that the
  BIMA HCN image contains typically 70\% of the single dish flux
  density.

  A 3~mm continuum image was produced as well by summing over the line
  free channels and both sidebands. The beam size of the image is
  15.5\arcsec$\times$10.5\arcsec\ (PA=--6\degr) and the rms is
  2.5~mJy/beam.

 \section{Results}
  \subsection{The 8 GHz continuum source associated with HD~37903}
  From the VLA continuum image\footnotemark, a weak, extended ($\sim
  40$\arcsec\ or 0.1~pc) source with an integrated flux density of
  $8.5\pm 1$~mJy is detected (Fig.~\ref{line+cont+c91a}).  The
  emission peaks (--10\arcsec, --12\arcsec) offset from HD~37903,
  corresponding to a projected distance of 0.035~pc from the exciting
  star. There is no evidence for continuum emission associated with
  the PDR.

  \footnotetext[5] {FITS images of the maps shown in this paper may be
    obtained at URL http://imagelib.ncsa.uiuc.edu}

  We assume that the observed continuum emission is due to optically
  thin free-free radiation from gas ionized by HD~37903. This is
  consistent with the low observed peak brightness temperature of
  0.12~K.  The evidence suggests that the stellar Lyman continuum flux
  is incident upon a nearby density enhancement.  We find for an
  electron temperature of 10000~K that the observed flux density is
  due to a region with an electron density of 200~\percc\ or an
  emission measure of 3000~${\rm cm}^{-6}\,{\rm pc}$. The Lyman
  continuum flux to maintain the ionization is
  $\log(N_C)=44.26$~s$^{-1}$, corresponding to ionization by a B2.5~V
  star \citep{pan73}. This value is in contrast to the observed
  optical spectral classification of B1.5~V \citep{rac68} which
  suggests either a density bounded \HII\ region or that a
  considerable fraction of Lyman continuum photons are absorbed in
  lower density more extended material not detected by the
  interferometer, thus lowering the spectral type estimate.

  There are, however, no molecular line features on our BIMA images
  associated with the \HII\ region though there does appear to be some
  associated molecular hydrogen just south of HD~37903 (see
  Fig.~\ref{overlay}). In addition, a streamer of \CNOA\ emission
  reaches from the \CII\ filament up to the \HII\ region.  We presume
  therefore that the ``nearby density enhancement'' contains rather
  little neutral gas (corresponding to a column density of at most
  $10^{21}$ \cmsq ) but that it should be observable in H$\alpha$ and
  other tracers of ionized gas.

  \subsection{Carbon recombination line image with the VLA}
  Images of the \CNOA\ emission averaged over the line are presented
  in Fig.~\ref{line+cont+c91a} with angular resolutions of 10 and
  20\arcsec. The carbon recombination line emission originates from a
  filament of roughly 3 arc minutes in length (0.4~pc) which extends
  from the east to the south of HD~37903. The projected distance from
  the exciting star varies between 0.13 and 0.17 parsec. Three
  emission maxima can be distinguished, positions 1, 2, and 3
  (Fig.~\ref{line+cont+c91a}).  On Fig.~\ref{line+cont+c91a}, also
  spectra averaged over 20\arcsec\ regions are shown centered on those
  maxima.  Table~\ref{line-parameters} lists the derived line
  parameters toward these positions.

  Based on the VLA data, the emission is resolved with 10 arc second
  resolution.  The width (perpendicular to the filament) of the \CNOA\ 
  emission is in the range 10--40\arcsec\ (0.02--0.08~pc) along the
  filament. There is no evidence for clumping on angular scales
  smaller than 10\arcsec, as had been earlier found for molecular
  hydrogen emission \citep{fie98}.  The 20\arcsec\ resolution image
  shows the emission with a higher dynamical range. In the remainder
  of the paper only images with this angular resolution are considered.

  As discussed in Sect.~\ref{sec-vla}, there is evidence for extended
  emission not detected by the VLA.  Figure~\ref{missing} shows the
  result of a comparison of the 100-m profile of \CNOA \ with our VLA
  results, both integrated over the whole emission area.  Some of the
  blue wing emission observed with the 100m is absent in the VLA data.
  Figures~\ref{overlay} and \ref{scuba} show integrated \CNOA \ images 
  with the total flux density fixed to the 100m value during the
  deconvolution.  The main difference with the image in
  Fig.~\ref{line+cont+c91a} is an additional halo of \CNOA\ around the
  clumps.  This halo is brighter toward the exciting star and the
  emission decreases sharply away from the star.

  Fig.~\ref{sumvlm} shows a position--velocity plot of the \CNOA\ 
  emission along the filament averaged over 40\arcsec\ perpendicular
  to the filament. Across clump 1 (Fig.~\ref{line+cont+c91a}), a
  velocity gradient of 7~\kms$\,{\rm pc^{-1}}$ is clearly observed,
  possibly reflecting rotation of the clump.  The remainder of the
  filament is quiescent with linewidths of $\sim 0.8$~\kms\ (see clump
  2 in Table~\ref{line-parameters}).  An overall view of the velocity
  structure is given in Fig.~\ref{cubes}, where a comparison of \CNOA\ 
  with the corresponding molecular line channel images is shown.

  The line observed in the direction of clump 2 is quite narrow ($\sim
  0.8$~\kms). The observed linewidth can be used to impose a stringent
  upper limit on the clump kinetic temperature \TKIN.  For purely
  thermal broadening, the linewidth $\Delta v_{th}$ (in \kms) will be
  given by:
  \begin{equation}
   \label{eq_dv}
  \Delta v_{th}^2 \, = \, 0.0458\, T_{\rm kin}/M({\rm AMU}),
  \end{equation} 
  where $M=12$ is the mass of the emitting species in AMU. From this,
  we derive an upper limit of 170~K. The line width toward clump 1 is
  larger, however, as discussed in the previous paragraph, it is
  dominated by the velocity gradient seen in Fig.~\ref{cubes} and {\it
    not} by an increase in the kinetic temperature.  Thus the limit
  which we derive for clump 2 may be valid along the whole filament.

  From the [\CII] 158~$\mu$m fine structure line intensity, a lower
  limit on the temperature of 138~K can be derived \citep{ste97}.
  Using the ratio of \CNOA\ and [O I] 63~$\mu$m emission,
  \cite{wyr97b} derived a temperature range of 100--300~K. Another
  lower limit on the temperature is the line intensity of the CO(7--6)
  line measured by \cite{jaf90}, which indicates a temperature of
  $>85$~K. In PDR models, the higher CO lines originate from about the
  same depth in the PDR as the carbon recombination line (i.\ e.\ the
  surface of the C$^+$/C/CO transition zone). These estimates are
  close to the upper limit inferred from the line width above. This
  agreement could suggest that the major line broadening agent is
  thermal and that the turbulent broadening is less than $\sim \, 0.4$
  \kms .  This value is in contrast to the situation in neighboring
  molecular layers (see Sect.~\ref{sec-molecules}) deeper in the PDR
  where the temperature is likely to be lower; without turbulent
  broadening the widths of the molecular lines would be considerably
  smaller than the \CNOA\ line widths but are observed to be similar
  to the \CNOA\ line width.  We conclude that the Alfv\`{e}n waves or
  other turbulent agents which give rise to non-thermal broadening in
  the molecular gas do not penetrate very much into the higher
  ionization degree layer associated with the \CNOA \ emission.

  Similar conditions may prevail in the reflection nebula NGC~7023
  where the 100-m observation of \CNOA \ (see Appendix) indicate a
  rather similar limit on the temperature.  Thus, we conclude that the
  temperature in the ionized carbon layers of PDRs excited by early
  B-stars with incident radiation fields similar to that estimated in
  NGC~2023 (smaller than $10^4$ times the average interstellar FUV
  field) is likely to be of order 200~K or lower.

  The observed spectral range covers the frequency of the sulfur (as
  well as Si and Mg) radio recombination lines. This line has been
  detected in several PDRs \citep{sil84} but not in the Orion Bar or
  NGC~2023. No S91$\alpha $ lines were detected in the VLA data by
  averaging over carbon line emission regions, in which the \CNOA\ 
  line was detected with a signal-to-noise ratio of 10.  Therefore the
  $2\sigma$ upper limit for the ratio of sulfur to carbon line
  intensities is 0.2, consistent with solar abundances
  \citep[C/H=3.6$\times 10^{-4}$ and S/H=2.1$\times 10^{-5}$]{gre96}
  and also with depletion of C by the factor observed in nearby
  diffuse clouds \citep[C/H=1.4$\times 10^{-4}$]{sof97} and solar
  abundance for S.  However, this limit rules out strong deviations
  from the C/S cosmic abundance ratio due to depletion of carbon on
  grains which was earlier found for S88B by \cite{sil84} and
  \cite{gar98}.

  \subsection{\CSFA \ line emission with the 100-m}

  Figure~\ref{c65a} shows the \CSFA\ carbon recombination line
  spectrum observed toward NGC~2023.  This observation is the highest
  frequency at which the carbon line in NGC~2023 has been detected to
  date.  Based on this detection it is possible to constrain the
  departure coefficients for principle quantum numbers $n$ between 65
  and 91; thus, in principle, the density and temperature in the PDR
  can be constrained.  The \CNOA \ VLA total flux corrected image was
  convolved to a 40\arcsec \ beam and a ratio of the integrated
  intensities $I_{91}/I_{65}$ of \CNOA \ and \CSFA\ of $2.4\pm 0.7$
  was estimated.  The ratio of departure coefficients $b_{91}/b_{65}$
  (see Eq.~7.3 of \cite{wyr97} for the case with no continuum) is:
  \begin{equation}
   b_{91}/b_{65} \, = \, \exp(18.28/T)\, (I_{91}\nu_{91})/(I_{65}\nu_{65}).
  \end{equation}
  We conclude that for a temperature $T$ of 150~K based on the above
  discussion, the measured intensity ratio implies $b_{91}/b_{65} =
  1\pm 0.3$.  This value compares with a value of 1.5--2 for
  $b_{91}/b_{65}$ expected on the basis of conventional hydrogenic
  recombination theory \citep{sal79} with an electron density in the
  range 3--30 \percc \ and temperature $\sim 150$~K.  Hence, we cannot
  explain this result by conventional hydrogenic recombination theory.
  As was the case for the Orion Bar \citep{wyr97a}, dielectronic
  recombination \citep{wal82b} seems most likely.

  \subsection{Molecular line emission images}
  \label{sec-molecules}

  Channel images of the \HCOP\ and HCN emission are shown in
  Fig.~\ref{cubes}, sampled with the same velocity grid as the \CNOA\ 
  emission. The brighter \HCOP\ emission is shown in greyscale with a
  complex, clumpy structure. All the bright \HCOP\ features are also
  observed in HCN. Differences between the two molecular transitions
  are discussed below.

  Figure~\ref{bima_spec} shows averaged spectra toward the two
  positions with strongest emission. The hyperfine satellites of HCN
  ($F=2-1,1-1,0-1$) are clearly detected with integrated intensity
  ratios of (1:0.44:0.26) at offset (52\arcsec,--43\arcsec) and
  (1:0.52:0.33) at offset (34\arcsec,--96\arcsec) from HD~37903,
  respectively.  These ratios are clearly different from the values
  expected in the optically thin LTE limit (1:0.6:0.2) and moreover the
  $F=1-1/2-1$ ratio is $<0.6$ and thus incompatible with the ratios
  expected in LTE at any optical depth.  Such anomalies in the HCN
  $J=1-0$ hyperfine satellite line ratios are well known
  \citep{gui81,wal82,gon93} and can be explained as being due to the
  combination of moderate opacity and line overlap. Indeed, the
  calculations of \cite[Fig.~2]{gui81} are able to reproduce our
  results toward the position (52\arcsec,--43\arcsec), where we
  detected the hyperfine satellites with brightness temperatures of
  (5.3~K, 2.2~K, 1.3~K), corresponding to $N({\rm HCN})\approx 5\times
  10^{12}$~\cmsq\ in their model with \TKIN=30~K and
  $n(\MOLH)=10^5$~\percc.

  Moderate opacity is also suggested by the fact that the line
  profiles of HCN and \HCOP\ agree. The northern clump (upper panel)
  shows a line width of 0.8~\kms, whereas the southern clump (lower
  panel) splits up into two velocity components separated by 1.6 \kms.
  This latter result suggests the presence of at least two clumps of
  higher density material with differing velocities immersed in the
  PDR.

  For the HCN and \HCOP(1--0) lines, the upper energies and critical
  densities are quite similar.  Hence changes in the relative
  intensities should reflect changes in the relative abundances of the
  two molecules. Fig.~\ref{ratio} shows the ratio of HCN and \HCOP\ 
  emission superimposed on the integrated \HCOP\ image.  Toward the
  emission peaks, the HCN/\HCOP\ ratio is reduced by a factor of two
  compared to the surroundings. In the southern clump, there is
  evidence for a gradient of the HCN/\HCOP\ ratio, decreasing from 0.8
  at the northern edge to 0.3 to the south.  \cite{fue93} have
  constructed chemical models which include HCN and \HCOP\ using
  densities and temperatures similar to NGC~2023.  In their models the
  \HCOP\ abundance decreases significantly for $A_v< 6$ due to the
  increasing electronic abundance. This effect might explain the
  increased HCN/\HCOP\ ratio toward the edges of the PDR.

  \subsection{3~mm continuum results}
  \label{sec-3mmcont}

  Within the observed mosaic, one resolved millimeter continuum source
  is detected at (--19\arcsec, --106\arcsec) offset from HD~37903
  (Fig.~\ref{scuba}). The deconvolved size of the emission is
  12\arcsec\ (0.028~pc) with an integrated flux density of
  36$\pm$7~mJy.  Based on the noise in our 8~GHz continuum image we
  can estimate an upper limit on free-free emission at 3~mm of 9~mJy,
  assuming optically thick emission from 8 to 85~GHz. A more realistic
  turnover from optically thick to thin emission at intermediate
  observing frequencies would lead to corresponding lower limits on
  free-free emission at 3~mm.  Thus, thermal dust emission must be the
  primary mechanism for the 3~mm continuum.  Using the equation for
  the corresponding mass given by \cite{mez90} with the dust cross
  section estimates by \cite{ren84} for dust with ice mantles, a dust
  opacity index of 2, and a temperature of 20~K, we find a clump mass
  of 6~\solmass.  The central column density of \MOLH \ is then
  $10^{24}$ \cmsq \ corresponding to a visual extinction of $\sim
  1000$~mag and an average molecular hydrogen density of
  $10^7$~\percc.

  Dust emission toward this position was first detected by
  \cite{lau96} with the SEST telescope at 1.3~mm; however with the
  lower angular resolution of 23\arcsec\ and a lower S/N no
  well-defined compact clump, as detected at 3~mm, was found.
  Fig.~\ref{scuba} shows a comparison of the 3~mm continuum emission
  with a 850~$\mu$m SCUBA image (G. Sandell 1999, priv.\ comm.) and
  the \CNOA\ and HCN line emission. The 3~mm source is coincident with
  compact 850~$\mu$m emission detected by SCUBA, confirming the
  interpretation that the source is caused by dust emission. Indeed it
  seems likely to be a cold ``prestellar core'', as the SCUBA
  detection is consistent with a temperature of 20~K. No IRAS source
  is at this position, however the IRAS images of this region are
  very confused. There seems to be a source associated with the core
  in a recent ISOCAM image of NGC~2023 (Nordh et al., in preparation),
  but no flux density estimates are available yet. The core is also
  coincident within the errors with star 3 of the \cite{dep90} NIR
  observations, who find this star to be the reddest object in the
  NGC~2023 star cluster with about 30~mag of visual extinction, assuming
  that the star is on the main sequence.  This
  extinction is still far smaller then the one derived for the
  millimeter core.  Hence, the star seems to lie in the foreground of
  the dense millimeter clump, but is still considerably embedded in
  the outer portion of the clump to account for the reddening. 
  Alternatively, the NIR emission might be due to scattered light from
  a more embedded source, since the emission seems to be resolved
  in the NIR images of \cite{fie98} and \cite{mcc99}.

  The 3~mm core has no prominent counterpart in HCN or \HCOP \ 
  emission. However, there is weak \HCOP \ emission towards the 3~mm
  continuum source with a linewidth of 0.8~\kms.  Combining this
  linewidth with the size estimate leads to a ``virial'' mass for the
  continuum source of roughly 2~\solmass. In addition, 30\arcsec\ to
  the NE of the 3~mm source, a prominent feature in the HCN, \HCOP\ 
  and the \CNOA \ images is observed with a velocity of 10.85~\kms\ 
  (Fig.~\ref{cubes}). The emission seems to originate from a small
  clump which appears to be on the surface of the 3~mm core. We will
  discuss this in more detail in Sect.~\ref{sec-model}.

 \section{Comparison of different PDR tracers}

  \subsection{Vibrationally excited molecular hydrogen emission}
  \label{sub-molh}
  By using observations of vibrationally excited molecular hydrogen
  in the NIR, angular resolutions of $\sim 1$\arcsec\ can be
  achieved in the photodissociated gas \citep{fie98,rou97}.  These
  lines originate from a thin cloud layer and are excited by the
  stellar UV radiation. At these surfaces carbon is ionized and can
  emit recombination lines. It is therefore of interest to compare
  \HVIB\ and \CNOA.

  Figure~\ref{overlay} shows an overlay of \CNOA\ emission with
  vibrationally excited \MOLH\ \citep{fie98}. The coordinate system of
  the \HVIB\ image was established using the position of the star
  HD~37903 and the scale and orientation given by \citep{fie98}. In
  addition, the positions of all the stars measured in previous NIR
  studies \citep{sel83,dep90} are plotted on top of the \HVIB\ image.
  The astrometry of the \HVIB\ image seems to agree more with the
  \cite{sel83} positions and we consider the offsets of up to
  8\arcsec\ of the \cite{dep90} positions to be due to a small stretch
  and rotation of their image.  The \CNOA\ clumps are systematically
  offset from the \HVIB\ filamentary emission, displaced away from the
  exciting star HD~37903.  This displacement suggests that the \HVIB\ 
  emission arises from limb brightening at the edge of the \CNOA\ 
  clumps. However, some of the \HVIB\ streamers are {\it not}
  associated with \CNOA\ emission. The offsets between the peak
  emission of \CNOA\ and \HVIB\ for the clumps referred to in
  Fig.~\ref{line+cont+c91a} are 10, 4, and 7\arcsec, respectively.
  This effect is shown in detail in the cross-cuts shown in
  Fig.~\ref{strips}. Stationary PDR models suggest a displacement
  between these components by $N_{\rm H}\approx 4\times
  10^{21}$~\cmsq\ \citep[$A_v\approx 2$]{hol97}. Hence, assuming
  edge-on geometry, gas with an atomic hydrogen density of
  6--14$\times 10^4$~\percc, could account for the observed offset
  (see the discussion in Sect.~\ref{sec-model}).

  \subsection{HCO$^+$ emission}

  The dense molecular cloud adjacent to NGC~2023 is revealed by the
  high \MOLH\ density tracer \HCOP\ ($n_{\rm cr}\sim
  10^{4.5}$~\percc). Figure~\ref{overlay} compares the carbon
  recombination line results with the BIMA \HCOP\ velocity averaged
  image. The \CNOA\ filament follows precisely the edge of the
  extended \HCOP\ emission. However, the two southern peaks of the
  \CNOA\ emission seem to be bracketed by the \HCOP\ emission. This
  effect can be seen most clearly in the lower panel of
  Fig.~\ref{strips}, where the cross-cut through the southernmost
  \CNOA\ peak reveals \HCOP\ in front of and behind \CNOA, while
  \HCOP\ is weak at the \CNOA\ peak position.

  The \HCOP\ emission can also be compared with the location of
  vibrationally excited \MOLH: low level \HCOP\ emission seems to
  trace the same material as the \HVIB\ emission in front of the
  \CNOA\ emission peaks indicating that some \HCOP\ can survive even
  in the outermost PDR layers. This effect was also seen by
  \cite{fue96} in NGC~7023. \HCOP\ can be formed in the radical
  zone of a PDR through reactions of O with $\rm CH_2^+$ and $\rm
  CH_3^+$ or reactions of $\rm C^+$ with OH \citep{ste95}.

 \section{PDR model results}
 \label{sec-model}
 To interpret the observed projected offsets between the carbon
 recombination line and vibrationally excited \MOLH\ emission
 (Sect.~\ref{sub-molh}), the intensities of these PDR tracers in a PDR
 model proposed by \cite{tie85} are calculated with $n_{\rm
   H}=10^5$~\percc, $G_0=10^4$, and $X$[C]=$1.4\times 10^{-4}$, where
 $G_0$ is the incident FUV field in units of the average flux of the
 interstellar medium. For an estimate of $G_0$ in NGC~2023 the results
 of \cite{ste97} were used. The physical and chemical conditions of
 the model shown in Fig.~\ref{n5g4a1} can account for the observed
 offsets; on the other hand, we cannot rule out deviations from an
 edge-on geometry.  Therefore the actual offsets between \CNOA\ and
 \HVIB\ might be larger with a resultant decrease in the density.  As
 can be seen in Fig.~\ref{n5g4a1}, the upper limits on the temperature
 in the \CII\ region from the measured line widths are consistent with
 the temperature predictions of the PDR model. The bulk of the \CNOA\ 
 emission originates from a depth in the PDR, where the temperature is
 100--200~K. The calculated \CNOA\ intensity is within a factor two
 equal to the observed values (Table~\ref{line-parameters}).

 In a clump of sufficiently high density, the positional offset
 between the molecular line and carbon line emission should become
 quite small. The models suggest that the difference in the depths in
 the PDR should be roughly $A_v\sim 3$.  This corresponds to $\sim$
 0.45/$n_{6}$ arc seconds at the distance of NGC~2023, where $n_{6}$
 is the \MOLH \ density in units of $10^6$ \percc. Thus the offsets
 will be difficult to measure at densities much above $10^5$ \percc.
 We do in fact observe a clump where there is rough coincidence
 between the peaks of \CNOA, HCN, and \HCOP. This object is associated
 with the 3~mm continuum source, discussed in Sect.~\ref{sec-3mmcont}.
 We conclude that this density must be significantly higher than
 $10^5$ \percc.

 \section{Discussion and Conclusions}

 The current results indicate that the carbon recombination line
 emission, tracing the ionized carbon content of the NGC~2023 PDR,
 originates from the edge of the dense molecular cloud as observed in
 the high density tracers HCN and \HCOP.  \CNOA \ emission bracketed
 between the molecular tracers and the exciting star is observed, as
 expected based on models of PDRs.  The \CNOA \ appears also to be
 offset relative to vibrationally excited \MOLH, consistent with an
 atomic hydrogen density of $10^5$ \percc.

 The linewidth estimates for \CNOA \ towards clump 2 place an upper
 limit of 170~K on the kinetic temperature in the ionized carbon
 layer. This value is in fact not much larger than {\it lower limits}
 on the temperature derived independently by \cite{ste97} and
 \cite{wyr97b}. We conclude that line broadening due to turbulent
 motions is of minor importance in the PDR and that the energy density
 in turbulence is small.  This finding may have consequences for the
 possibility of stars forming within the PDR.

 An important result of our BIMA observations is the detection of the
 compact 3~mm continuum source. There is no IRAS source associated
 with it, but it is detected as a compact dust core at 850~$\mu$m (G.
 Sandell 1999, priv.\ comm.).  Furthermore, there is little sign of
 its presence in molecular lines.  These facts argue that the core is
 cold and therefore starless. Molecules seem to be frozen-out onto
 grain surfaces, since no prominent molecular line emission is
 detected in the direction of the core.

 The findings suggest that the dust core detected by BIMA is in the
 process of collapsing to form a protostar. Observations in other
 tracers can be important in order to test whether the absence of
 associated HCN and \HCOP \ is a general phenomenon or is confined to
 those two species. The detection of molecular lines associated with
 the object would considerably help in understanding its evolutionary
 state.  From the density estimate of $10^7$ \percc \ inferred from
 the continuum results, we derive a free--fall time of $10^4$ years.
 This time is an order of magnitude less than the timescale expected
 for evolution of the PDR taking a size scale of 0.04 parsec and a
 velocity of 0.5~\kms .  A possible scenario is that the dust core
 formed within the PDR with the formation triggered by the higher
 pressures resulting from the PDR heating. On the other hand, there is
 no evidence for interaction of the FUV radiation field with the
 BIMA/SCUBA core in the form of \CNOA \ emission.  However, the
 radiation field does seem to be interacting with the clump to the NE
 observed in \HCOP \ and HCN and it seems unlikely that the two clumps
 are adjacent only due to a chance projection.

  \acknowledgments
  We would like to thank Jean-Louis Lemaire for providing the \HVIB\ 
  image of NGC~2023 and Goeran Sandell for providing the SCUBA image
  of NGC~2023 prior to publication.  FW is supported by the National
  Science Foundation under Grant No. 96-13716. The research of CMW is
  partially supported by the ASI grant ARS-98-116 and by the MURST
  project ``Dust and Molecules in Astrophysical Environments''. The
  National Radio Astronomy Observatory is a facility of the National
  Science Foundation operated under cooperative agreement by
  Associated Universities, Inc.

 \appendix

 \section{\CNOA\ Effelsberg 100m observations}

 We have searched for emission in the \CNOA\ radio recombination line
 towards several other star forming regions harboring PDRs using the
 Effelsberg 100m telescope. A more detailed description of the
 observations is given in \cite{wyr97}. The observations were carried
 out in July and August 1996. Table~\ref{carbon_sources} gives a list
 of the observed sources and Fig.~\ref{fig-app} shows the observed
 lines. The frequency of the observed \CNOA\ lines is 8589.104~MHz
 with a beamsize of 82\arcsec. Pointing was checked through
 observation on NGC~7027, W3(OH), and 3C161.  Observations of NGC~7027
 were used to establish the brightness temperature scale assuming a
 continuum flux density of 6.25~Jy and a telescope sensitivity of
 2.7~K/Jy.  DR~21 was observed in position switching mode and all
 other sources in frequency switching mode. The spectrometer was an
 autocorrelator yielding a spectral resolution of 0.2~\kms\ in case of
 NGC~7023 and 0.4~\kms\ for the other sources.  The average
 integration time on a source was 7~hours.  In
 Table~\ref{carbon-line-parameters} the results of Gaussian fits to
 the observed spectra are listed.  In case of the three non-detections
 the 1$\sigma$ level of the noise in the observed spectra is given. An
 important result to note is the extremely narrow \CNOA\ line in
 NGC~7023. The linewidth is clearly smaller than the widths of \HVIB\ 
 and \CII\ 158~$\mu$m \citep{lem99,ger98}, indicating that the latter
 emissions originate from a hotter surface layer of the PDR.  Using
 Eq.~\ref{eq_dv} we estimate an upper limit of the temperature in the
 \CII\ zone of 190~K. This value is quite similar to the limit we
 derived for NGC~2023.


\clearpage

\bibliography{}


\clearpage

\begin{table}
\centering
\caption[vla parameters]
        {VLA Observing Parameters}
\vspace*{2mm}
\begin{tabular}{cc}
\hline
\hline
Rest Frequency (\CNOA ) &  8.589104~GHz   \\
Total Bandwidth         &  1.5625 MHz  \\
Number of Channels      &  255    \\
Channel Separation      &  6.1 KHz (0.21~km s$^{-1}$)  \\
Synthesized Beam FWHM   &   11.3 x 9.3~arcsec, PA=--3\degr \\
Tapered Beam FWHM       &   19.6 x 18.1~arcsec PA=1\degr\\
Primary Beam FWHM       &    5.2~arcmin \\
Phase Center            &  $\alpha_{2000}$= 5:41:38.4
                           $\delta_{2000}$=-2:16:32   \\
Time On Source          &  9.7 hours    \\
\hline 
\end{tabular}
\label{line parameters}
\end{table}

\clearpage

\begin{table}
\centering
\caption[line-parameters]
        {Line parameters of the averaged spectra shown 
         in Fig.~\ref{line+cont+c91a}.}
\vspace*{2mm}
\label{line-parameters}
\begin{tabular}{ccccc}
\hline
 clump \# & offsets  & $T_{\rm L}$ & $v_{\rm LSR}$ & $\Delta v$  \\
          & (arcsec) &    (K)      &  (\kms)       & (\kms)      \\
\hline
 NGC2023:CII 1 & (--7, --88) & 0.36$\pm$0.05 & 10.31$\pm$0.05 & 1.3$\pm$0.1 \\
 NGC2023:CII 2 & (25, --64)  & 0.49$\pm$0.06 &  9.97$\pm$0.03 & 0.8$\pm$0.1 \\
 NGC2023:CII 3 & (56, --7)   & 0.16$\pm$0.05 & 10.0$\pm$0.1   & 1.9$\pm$0.3 \\ 
\hline 
\end{tabular}
\end{table}

\clearpage

\begin{table}
\centering
\caption{Sources observed in the \CNOA\ carbon recombination line. }
\label{carbon_sources}
\vspace*{2mm}
\begin{tabular}{lcc}
\hline
Source & $\alpha$(2000) & $\delta$(2000) \\
       & h m s          & \arcdeg\phn\arcmin\phn\arcsec \\
\hline
IC 63        & 00 59 01.4 & \phs60 53 17.604 \\
RMC (IR6306) & 06 33 16.1 & \phs04 34 56.788 \\
S 106        & 20 27 26.6 & \phs37 22 47.872 \\
NGC 7023     & 21 01 37.2 & \phs68 09 46.587 \\
DR 21        & 20 39 01.0 & \phs42 19 29.855 \\ 
DR 21 (OH)   & 20 39 00.7 & \phs42 22 32.841 \\
\hline
\end{tabular}
\end{table}

\clearpage

\begin{table}
\centering
\caption{\CNOA\ line parameters. For non-detection the 1$\sigma$ noise
         level in the spectrum is given.}
\vspace*{2mm}
\begin{tabular}{lrcccc}
\hline
Source & Offsets & $\int T_{\rm MB}dv$ & $T_{\rm MB}$ & $v_{\rm LSR}$ & 
         FWHM \\
       &  (arcsec) &   (K$\,$km$\,$s$^{-1}$) & (K)  & (km$\,$s$^{-1}$) &
       (km$\,$s$^{-1}$) \\
\hline
IC 63         &            &      & 0.01 &      &      \\
RMC (IR6309)  &            &      & 0.01 &      &      \\
S 106         &  (0, 0)    & 0.69$\pm$0.03 & 0.07$\pm$0.02 & 
              -4.0$\pm$0.2 & 8.9$\pm$0.5  \\
S 106         & (30, --30) & 0.40$\pm$0.04 & 0.04$\pm$0.02 & 
              -3.5$\pm$0.2 & 9.8$\pm$1.1  \\
NGC 7023      & (--20, 30) & 0.11$\pm$0.01 & 0.12$\pm$0.02 &  
               2.3$\pm$0.05 & 0.88$\pm$0.08 \\
NGC 7023      &  (0, 0)    & 0.05$\pm$0.02 & 0.07$\pm$0.02 &  
               2.6$\pm$0.12 & 0.8$\pm$0.2  \\
DR 21         &            & 0.76$\pm$0.08 & 0.07$\pm$0.02 & 
              -5.1$\pm$0.4 & 10$\pm$1   \\
DR 21 (OH)    &            &      & 0.01 &      &      \\
\hline 
\end{tabular}
\label{carbon-line-parameters}
\end{table}


\clearpage

\figcaption[missing.eps]
  { 
    Integrated emission of Effelsberg 100m measurements of the \CNOA\ 
    carbon recombination line (80\arcsec\ beam) compared with the
    integral over the VLA observation (grey histogram).
  \label{missing}
  }

\figcaption[show_allvla.eps]
  { 
    Radio continuum (greyscale) and \CNOA\ carbon recombination line
    emission (contours) toward NGC~2023 observed with the VLA. Line
    emission is averaged from $v_{\rm LSR}$ 9.6--11.1~\kms. The left
    panel shows natural weighting images ($\sim$ 10\arcsec\ beam) and
    the middle panel smoothed images ($\sim$ 20\arcsec\ beam) using a
    uv-taper to emphasize emission on larger scales. Contours and
    greyscales start at 3$\sigma$ and advance in 2$\sigma$ steps.  For
    the line emission the -3$\sigma$ dashed contour is also shown. The
    noise in the continuum and \CNOA\ images in both panels are,
    respectively, 0.085, 0.41, 0.13 and 0.45~mJy/beam. The right
    panels show spectra of the \CNOA\ line averaged over 20\arcsec\
    regions centered on the clumps labeled in the middle panel.
    Parameters of the Gaussian fits (dotted lines) are given in
    Table~\ref{line-parameters}.
  \label{line+cont+c91a}
  }

\figcaption[manual_align_hco+.eps]
  {  
    Comparison of the locations of \CNOA\ emission (thick contours,
    maximum entropy deconvolution, cf.\ Sect.~\ref{sec-vla}) with
    those of vibrationally excited \MOLH\ (Field et al.\ 1998, thin
    contours) and \HCOP\ (greyscale).  The \CNOA\ emission is averaged
    from 9.6--11.1~\kms\ and the contours are 3,5,7$\dots \times$
    0.4~mJy/beam (19.6\arcsec $\times$18.1\arcsec\ beam). The \HCOP\ emission
    is averaged from 8.53--11.5~\kms\ and the contours are
    3,6,9$\dots \times$ 0.1~Jy/beam (11.2\arcsec $\times$8.1\arcsec\ 
    beam). As a check of the astrometry of the NIR image, the
    positions of the NIR sources (Sellgren 1983; DePoy et al.\ 1990;
    see corresponding markers) are shown. For the \HVIB\ and the
    \HCOP\ data the boundaries of the observations are outlined.
  \label{overlay}
  }

\figcaption[scuba.eps]
  { 
    Comparison of the locations of \CNOA\ emission (same as in
    Fig.\ref{overlay}) with those of 850~$\mu$m dust emission
    (greyscale, G. Sandell 1999, priv.\ comm.), HCN (dashed contours)
    and 3~mm continuum emission (thick, dotted contours). The HCN 
    emission is averaged from 8.81--11.8~\kms\ and the contours are
    3,5,7$\dots \times$ 0.15~Jy/beam (15.8\arcsec $\times$10.7\arcsec\ 
    beam).The contours of the 3mm continuum image are
    3,5,7$\dots \times$ 2.5~mJy/beam (15.5\arcsec $\times$10.5\arcsec\ 
    beam).
  \label{scuba}
  }

\figcaption[show_sumvlm.eps]
  { 
    Position--velocity plot along the \CNOA\ filament (PA= 51\degr).
    The solid lines denote the positions of the three clumps labeled
    in Fig.~\ref{line+cont+c91a}. The \CNOA\ emission is averaged over
    40\arcsec\ perpendicular to the filament and the contours are
    3,5,7$\dots \times$ 0.7~mJy/beam (19.6\arcsec $\times$18.1\arcsec\ 
    beam).
  \label{sumvlm}
  }

\figcaption[cubes.eps]
  { 
    Channel images of the \CNOA\ emission (thick contours,
    3.2~mJy/beam spacings, 19\arcsec\ beam) and the HCN emission (thin
    contours) on top of the \HCOP\ emission shown in greyscale (both
    molecular lines with 1~Jy/beam spacings, 15.8 x 10.7\arcsec\ 
    resolution, observed with BIMA). The filled square marks the
    position of the 3~mm continuum source.
  \label{cubes}
  }

\figcaption[n2023-c65a.eps]
  { 
    Effelsberg 100m spectrum of the \CSFA\ carbon recombination
    line observed at an offset position of (0'',-80'') from HD~37903.
  \label{c65a}
  }

\figcaption[bima_spec.eps]
  { 
    HCN and \HCOP\ (grey-filled histogram) spectra averaged over
    20\arcsec\ regions centered on the offset positions
    (52\arcsec,--43\arcsec) and (34\arcsec,--96\arcsec). The HCN
    hyperfine satellite lines (at --7.1 and 4.9~\kms) are detected.
  \label{bima_spec}
  }

\figcaption[ratio.eps]
  { 
    Ratio of HCN and \HCOP\ emission (greyscale, steps of 0.1 from 0.3
    to 0.8) overlaid on the integrated \HCOP\ emission (contours:
    30\%,50\%,70\%$\dots$ of the peak flux of 4.45~Jy/beam (15.8\arcsec $\times$10.7\arcsec\ 
    beam)).
  \label{ratio}
  }

\figcaption[strips.eps]
  { 
    Cross-cuts through the \CNOA\ clumps in
    \CNOA, continuum, \HVIB, \HCOP\ and HCN.  In each panel the
    orientation of the cross-cut and the offsets of the clumps from HD 37903
    are indicated. 
  \label{strips}
  }

\figcaption[do_graph_n5g4a1.eps]
  { 
    Model results for an homogeneous edge-on PDR (Tielens \&
    Hollenbach 1985) having $n_{\rm H}=10^5$~\percc, $G_0=10^4$, and
    $X$[C]=$1.4\times 10^{-4}$.
  \label{n5g4a1}
  }

\figcaption[newcarbon.eps]
  { 
    Carbon \CNOA\ recombination lines towards several PDRs observed
    with the Effelsberg 100m telescope. The broad feature in the DR21
    spectrum at 20~\kms\ is the helium line from the prominent \HII\ 
    region.
  \label{fig-app}
  }


\begin{figure}
  \epsfysize=\textwidth
  \rotatebox{-90}{\epsfbox{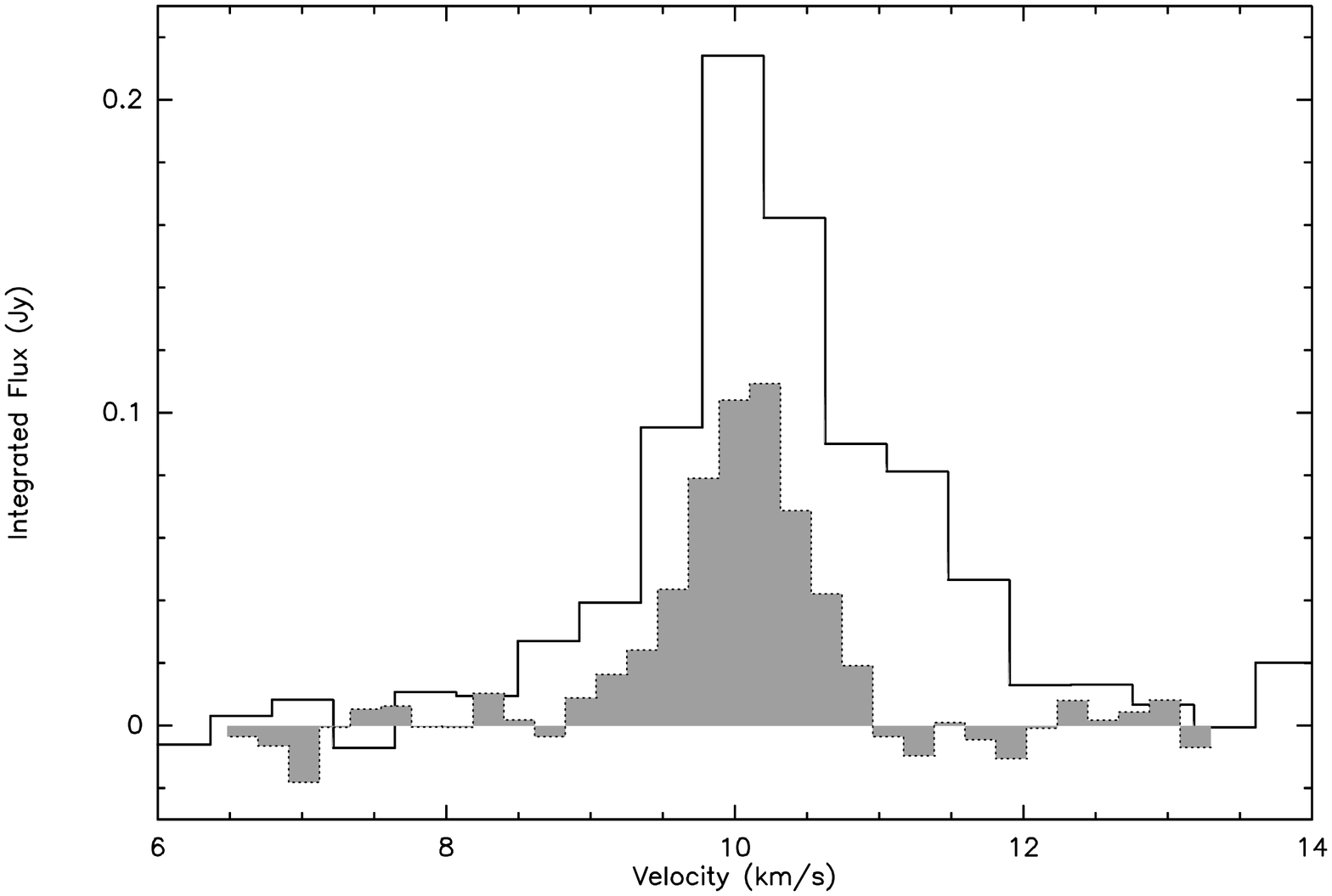}}
\end{figure}

\begin{figure}
  \epsfysize=\textheight
  {\epsfbox{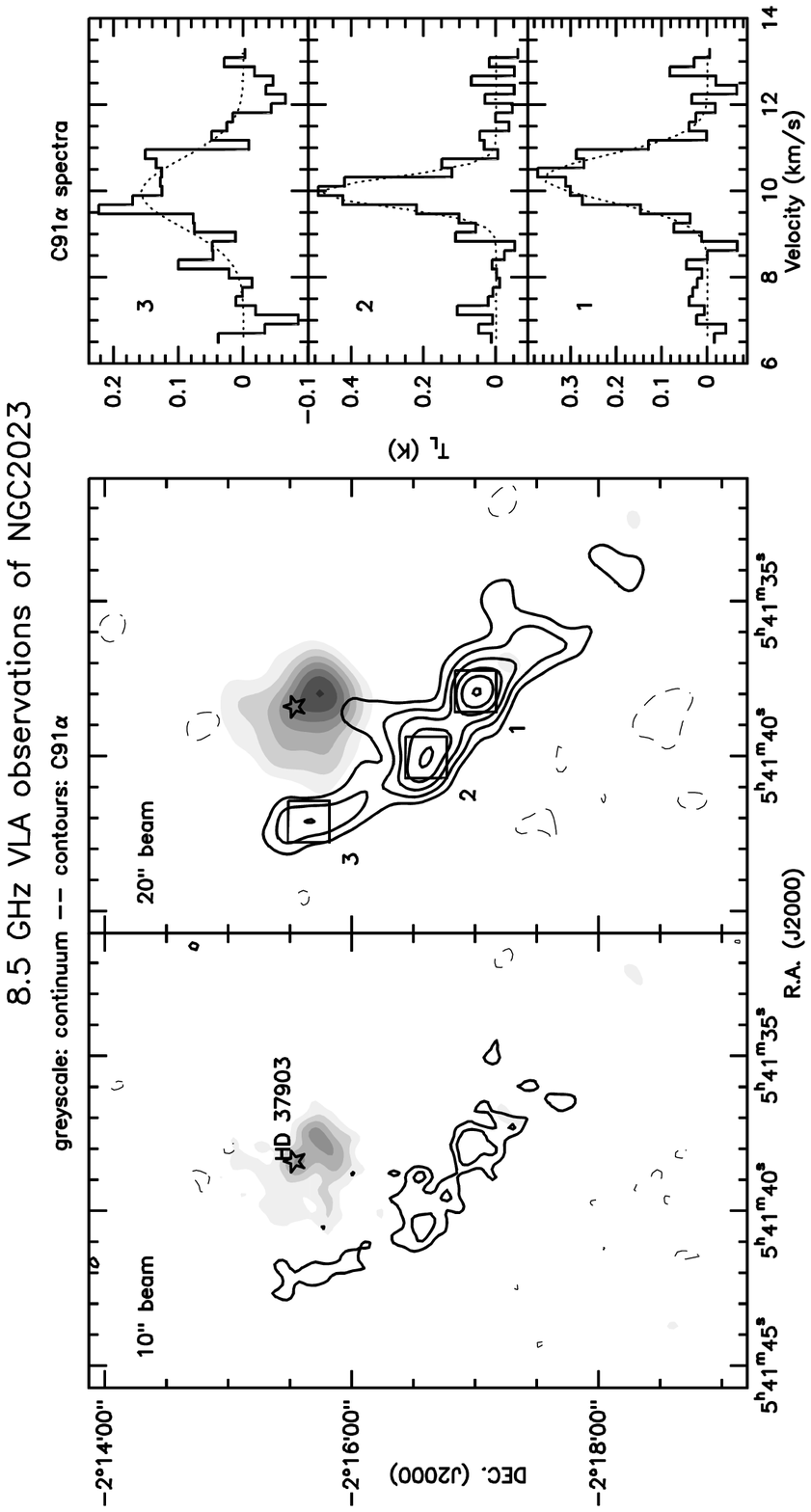}}
\end{figure}

\begin{figure}
  \epsfysize=\textwidth
  \rotatebox{-90}{\epsfbox{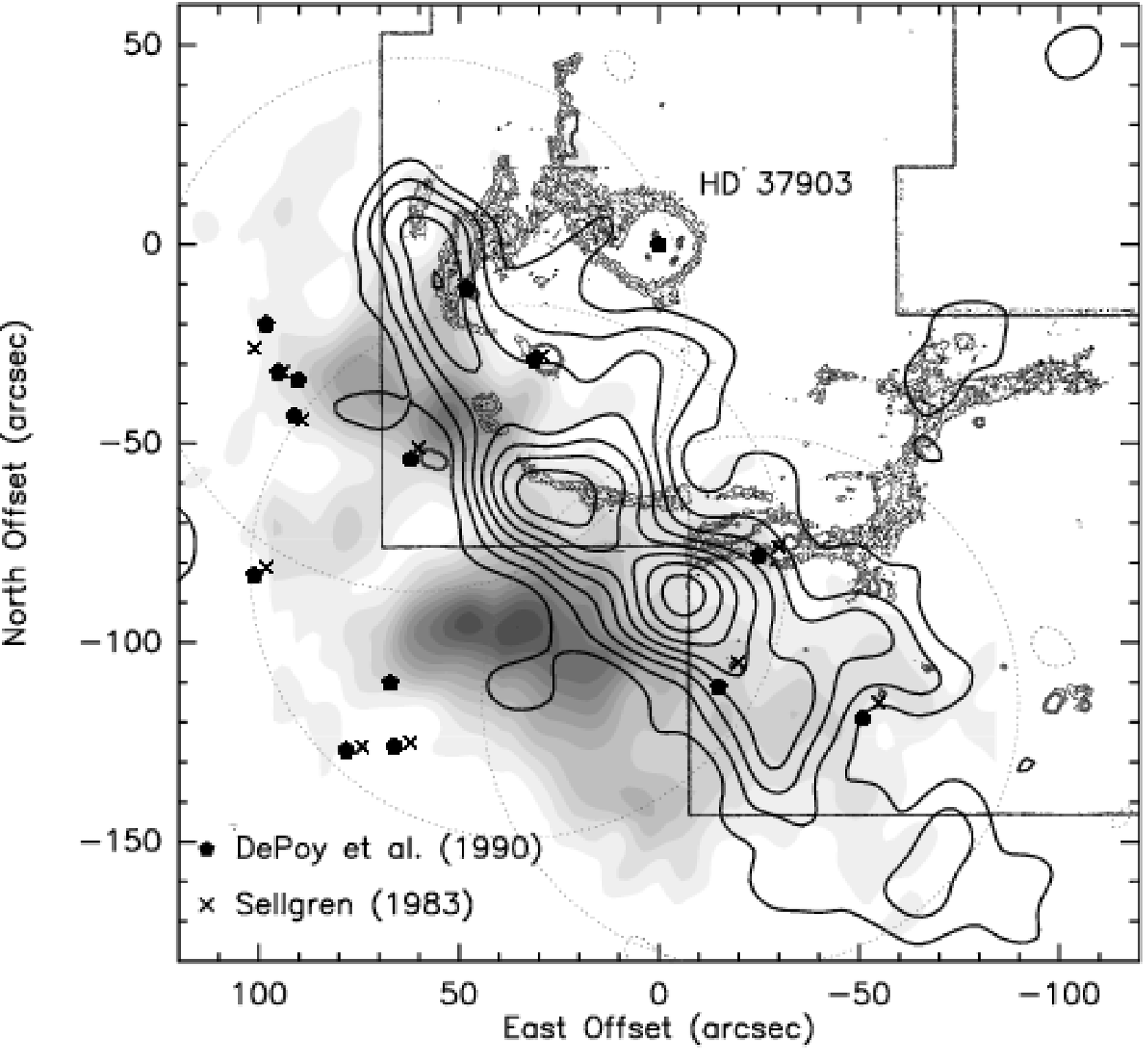}}
\end{figure}

\begin{figure}
  \epsfysize=\textwidth
  \rotatebox{-90}{\epsfbox{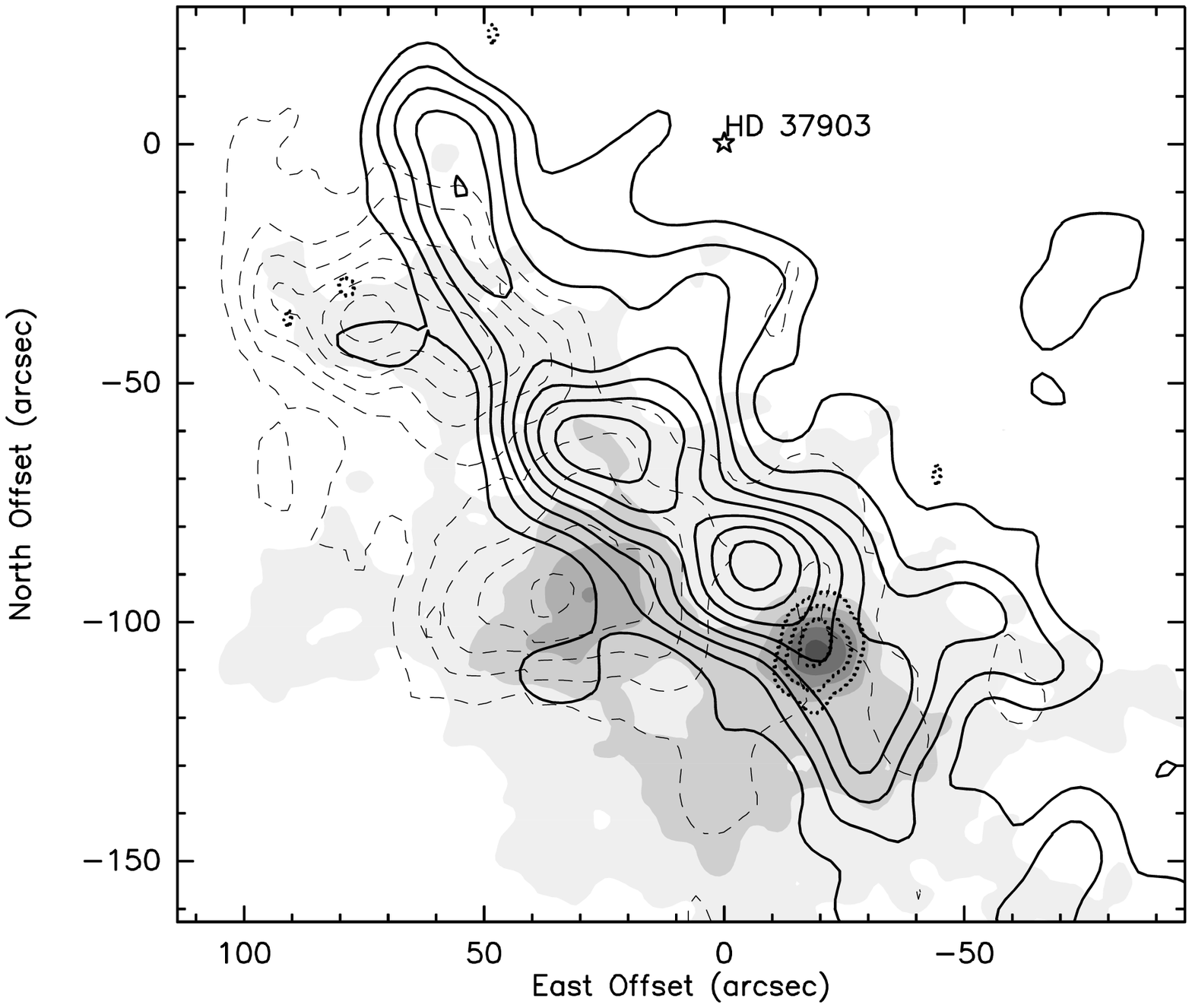}}
\end{figure}

\begin{figure}
  \epsfysize=\textwidth
  \rotatebox{-90}{\epsfbox{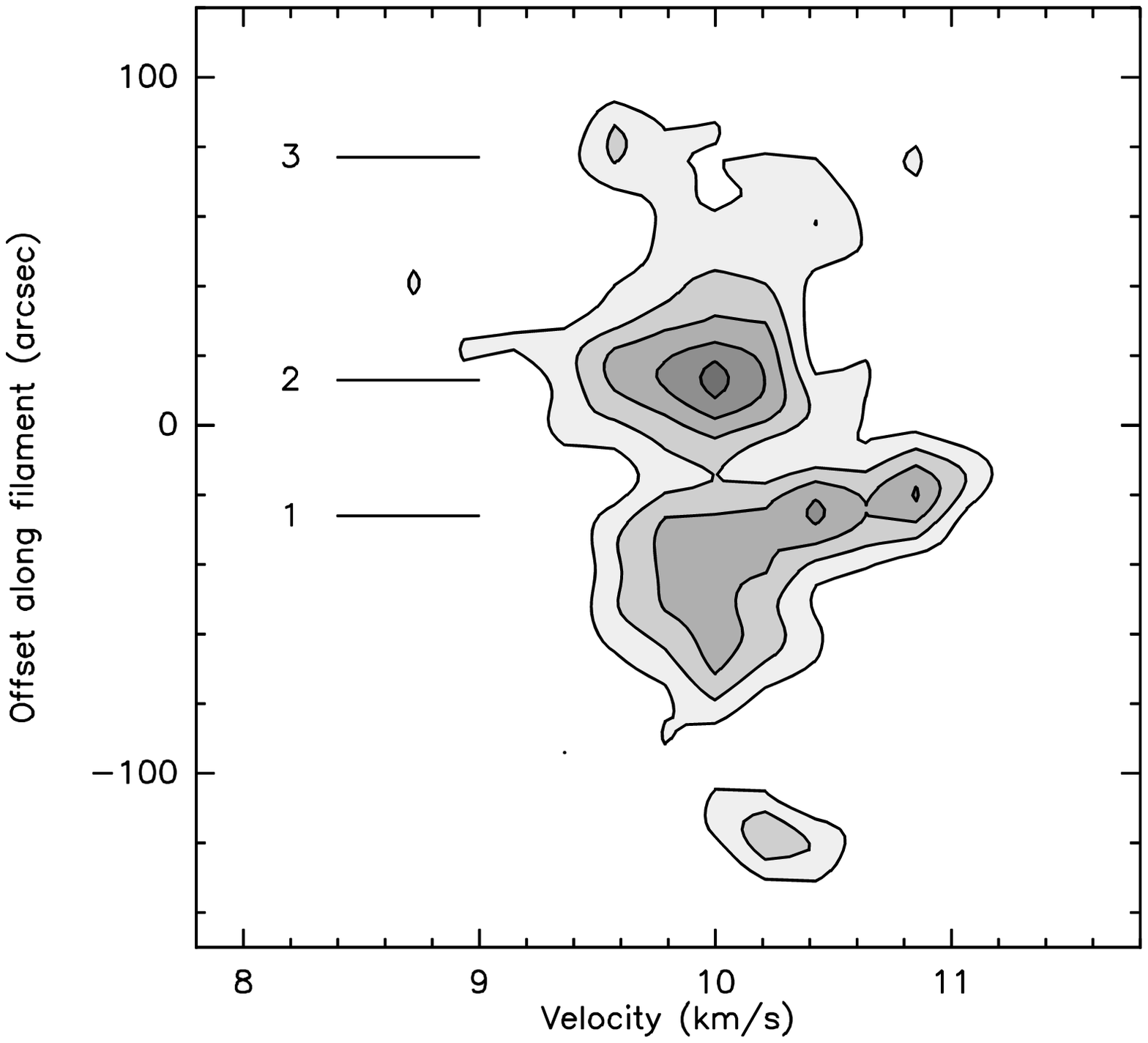}}
\end{figure}

\begin{figure}
  \epsfysize=\textwidth
  \rotatebox{-90}{\epsfbox{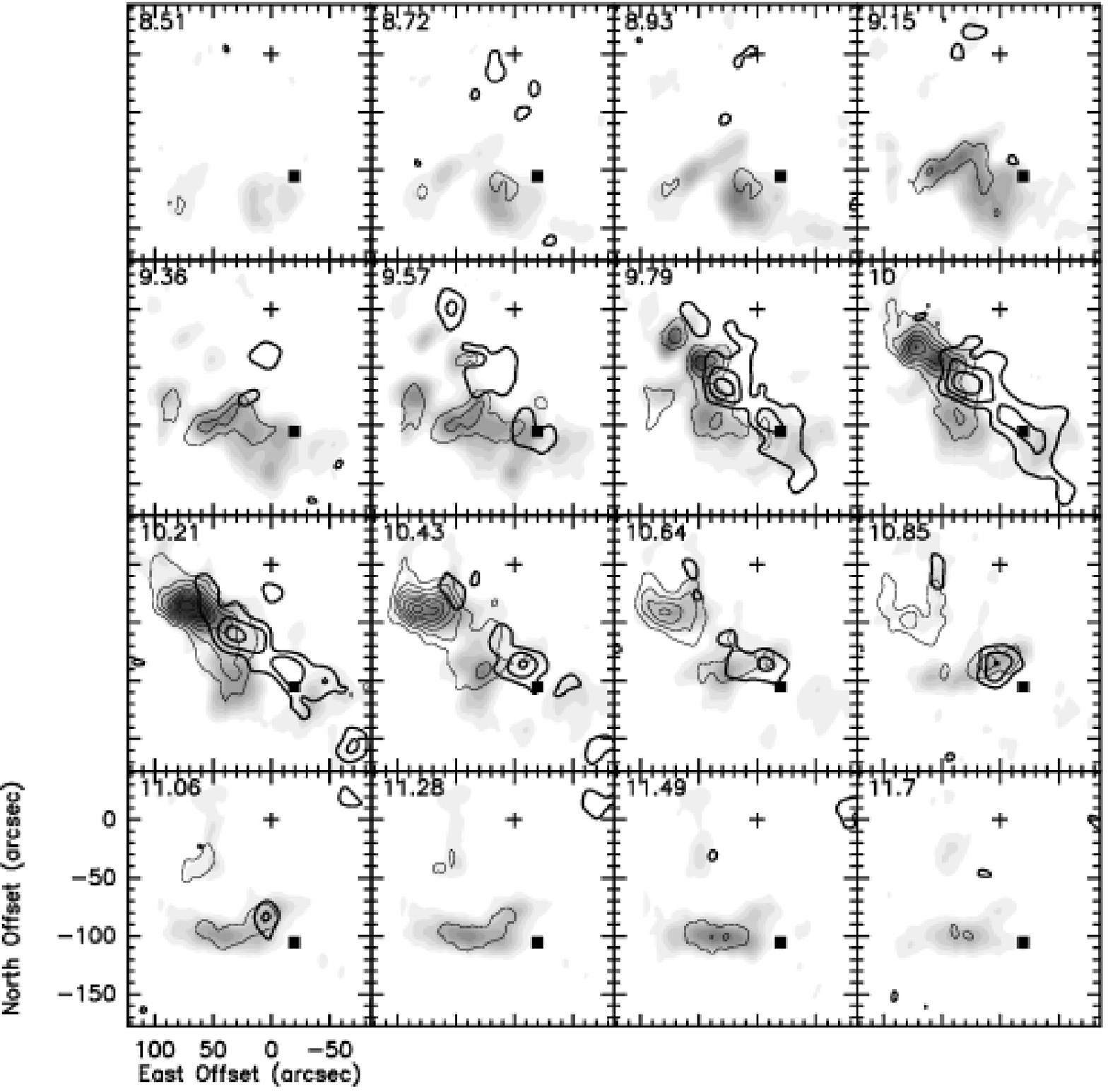}}
\end{figure}

\begin{figure}
  \epsfysize=\textwidth
  \rotatebox{-90}{\epsfbox{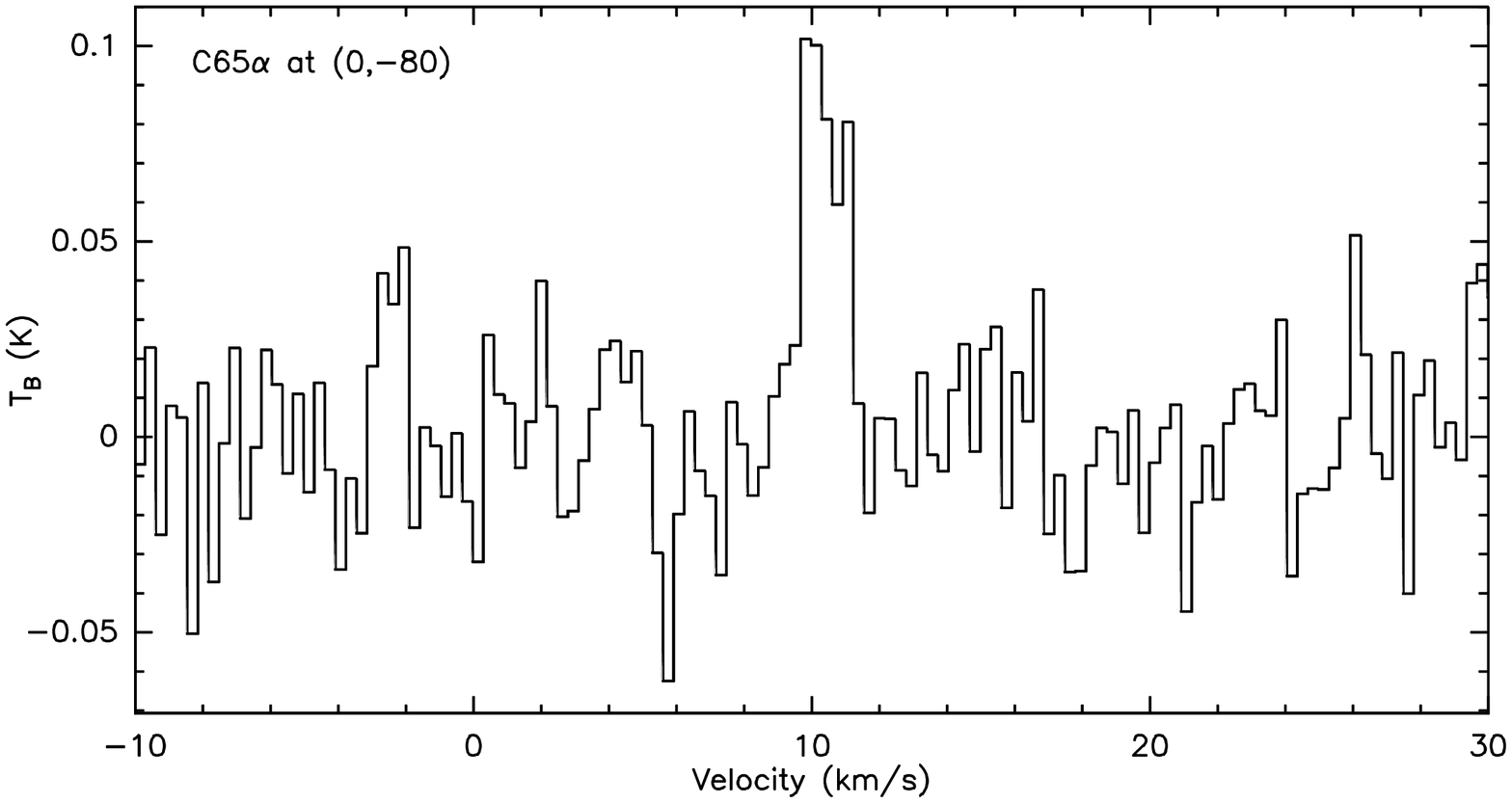}}
\end{figure}

\begin{figure}
  \epsfysize=\textwidth
  \rotatebox{-90}{\epsfbox{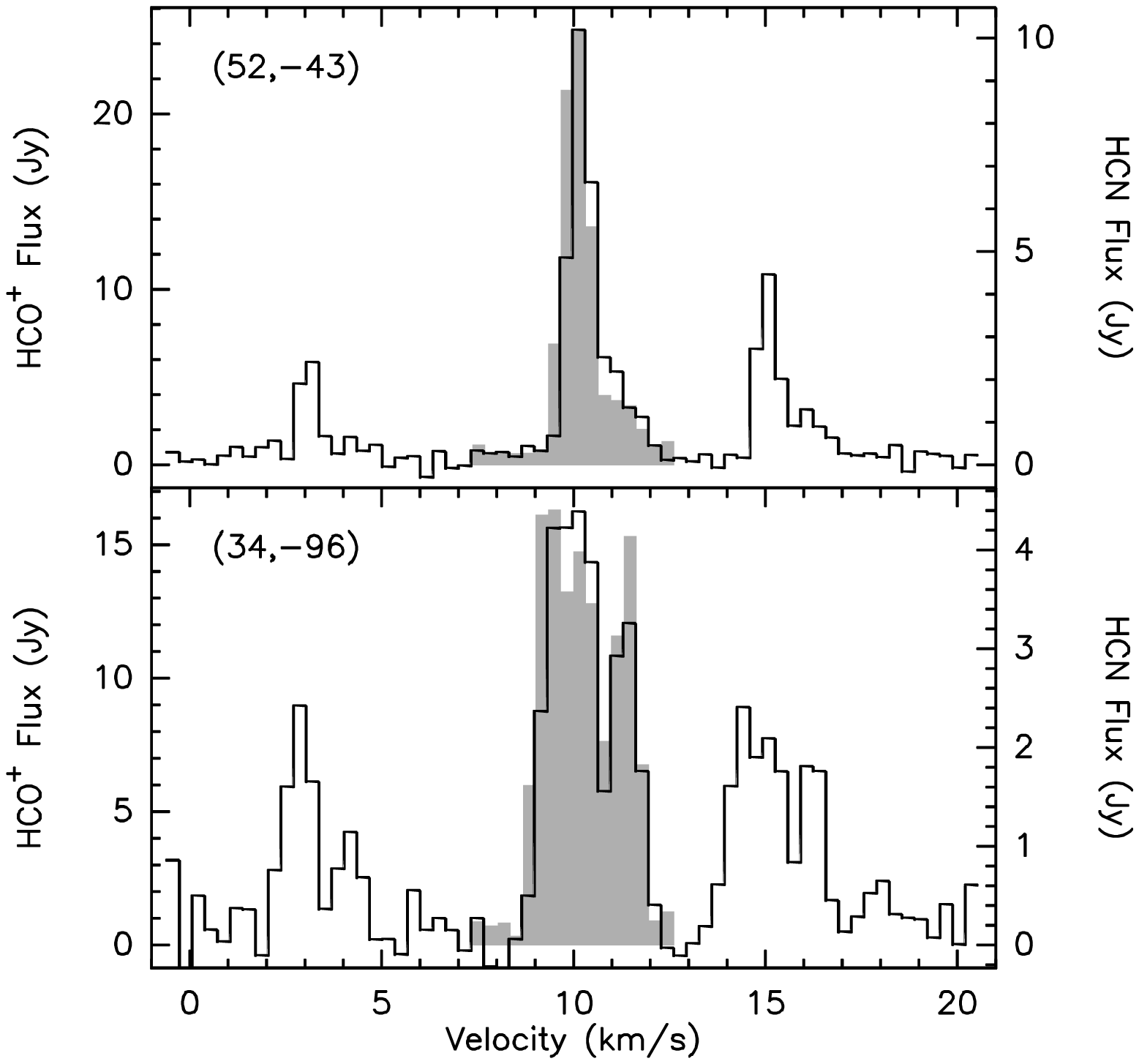}}
\end{figure}

\begin{figure}
  \epsfysize=\textwidth
  \rotatebox{-90}{\epsfbox{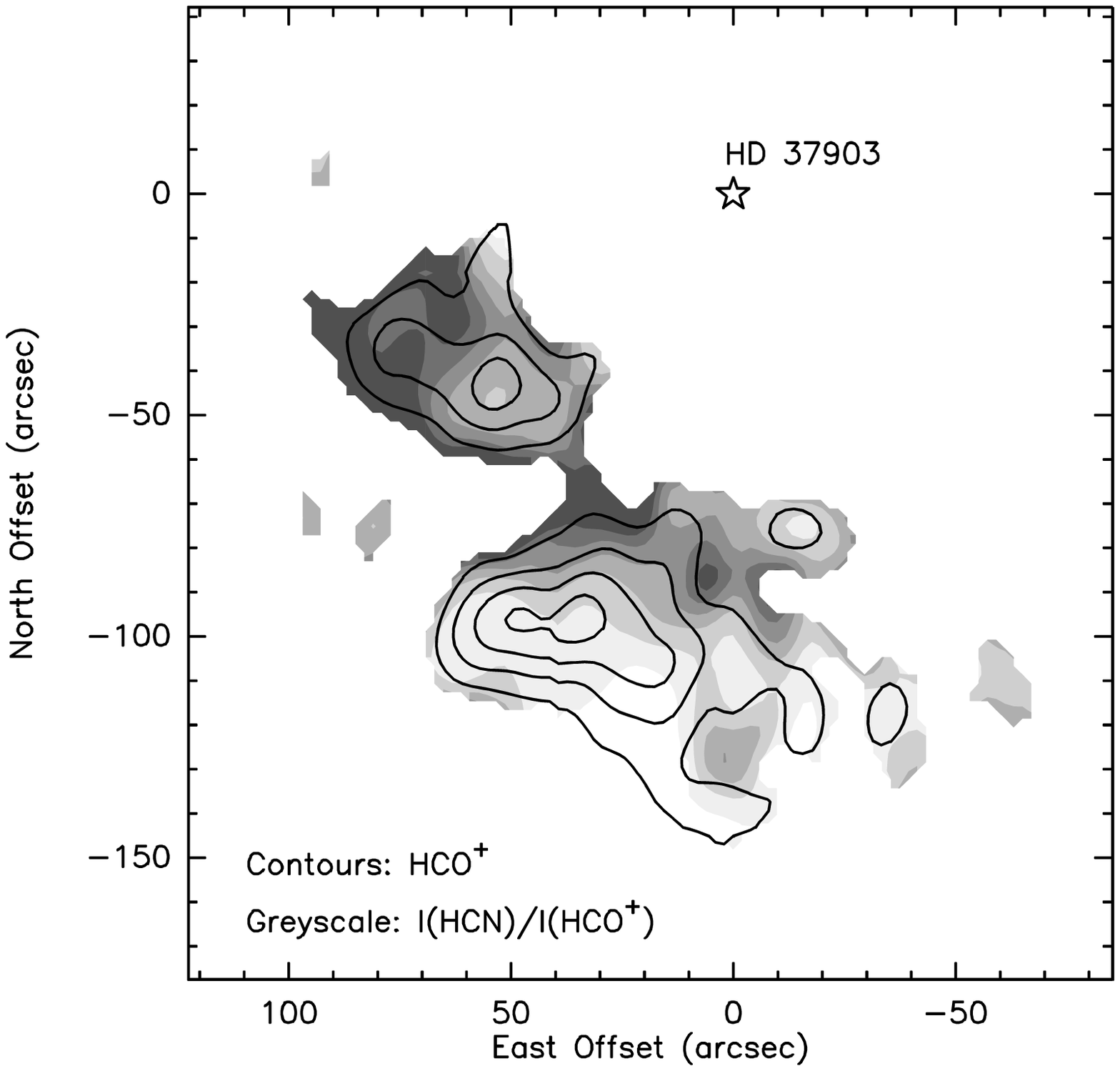}}
\end{figure}

\begin{figure}
  \epsfysize=\textheight
  {\epsfbox{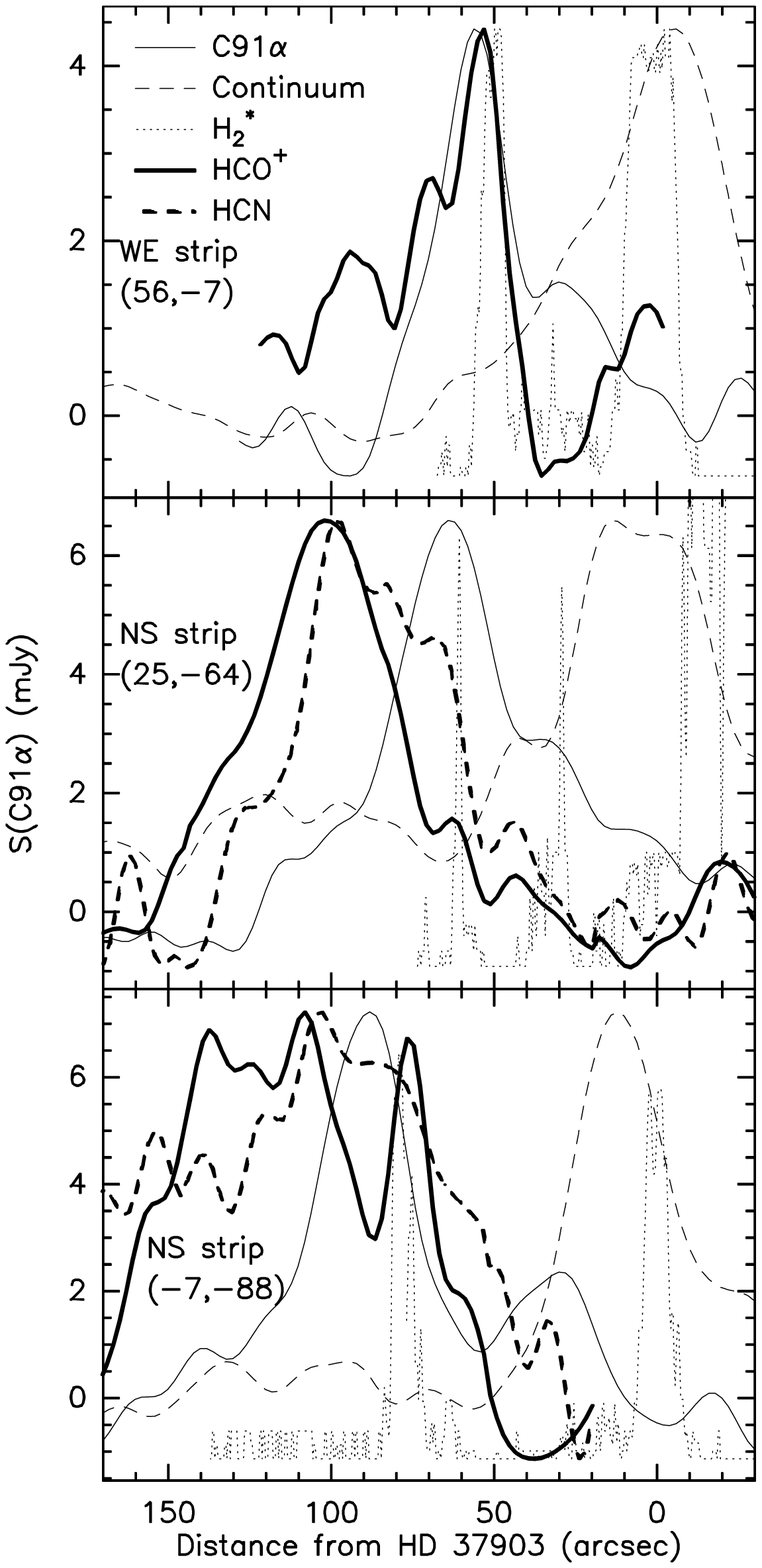}}
\end{figure}

\begin{figure}
  \epsfysize=\textwidth
  \rotatebox{-90}{\epsfbox{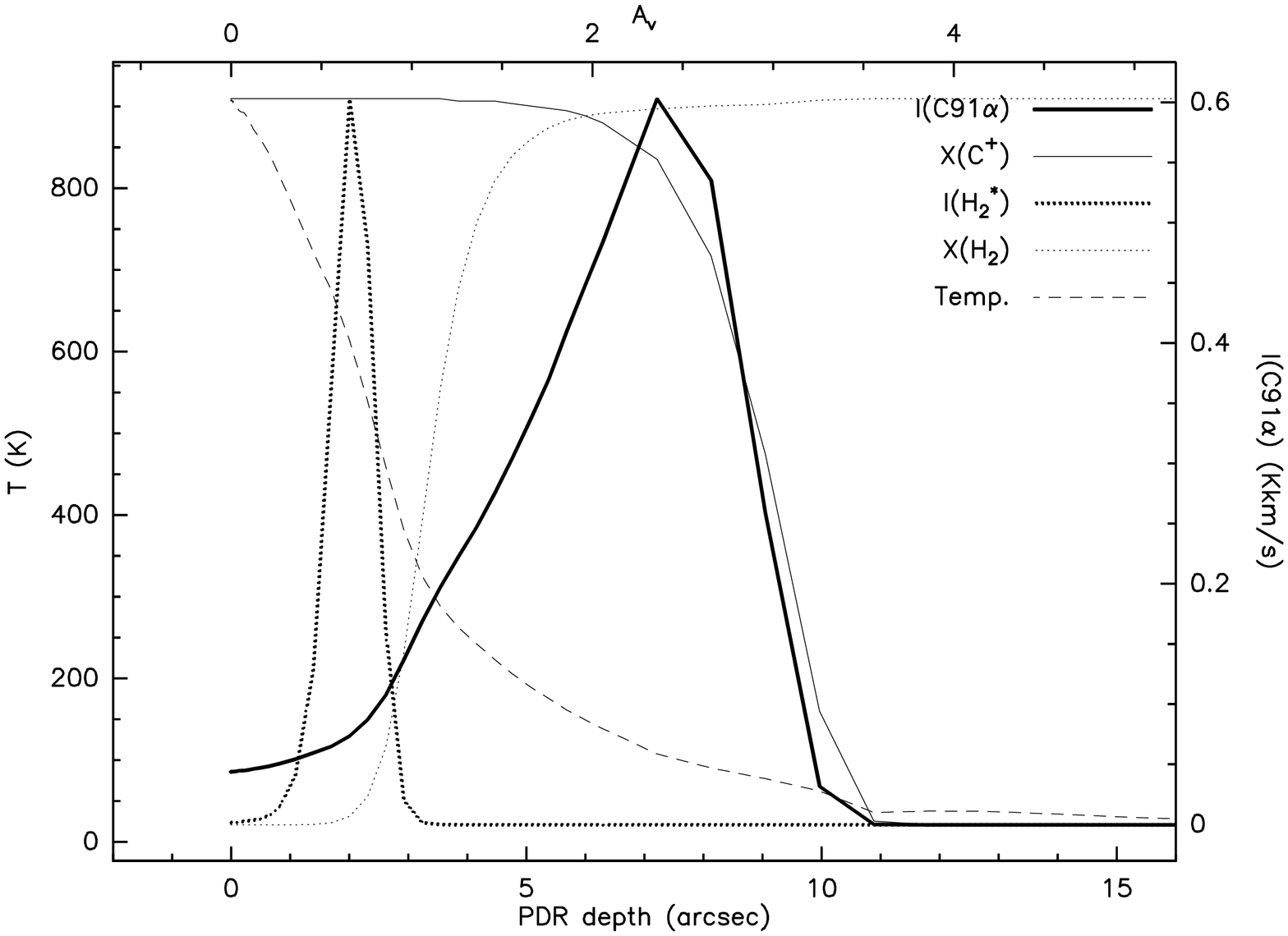}}
\end{figure}

\begin{figure}
  \epsfysize=\textheight
  {\epsfbox{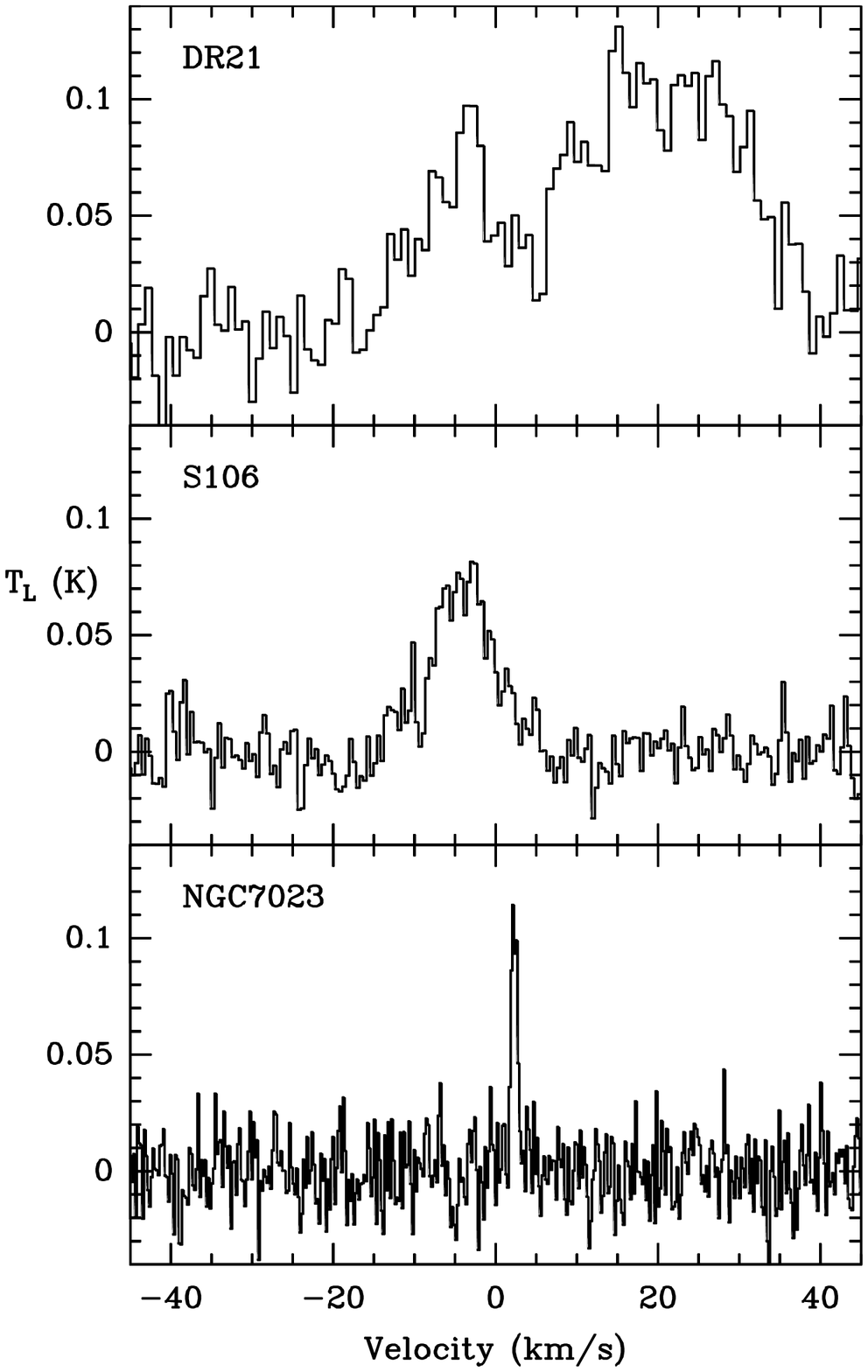}}
\end{figure}

\end{document}